\documentclass[a4paper, 12pt]{article}
\usepackage[dvips]{graphicx}

\def\ltap{\ \raise.3ex\hbox{$<$\kern-.75em\lower1ex\hbox{$\sim$}}\ }
\def\gtap{\ \raise.3ex\hbox{$>$\kern-.75em\lower1ex\hbox{$\sim$}}\ }

\begin{document}

\title{Chiral symmetry breaking and vacuum polarization in a bag}
\author{S.~Yasui\thanks{yasui@th.phys.titech.ac.jp} \\
\normalsize Physics Department, Tokyo Institute of Technology,\\
\normalsize Ookayama 2-12-1, Meguro, Tokyo 152-8551, Japan \\}
\maketitle

\begin{abstract}
We study the effects of a finite quark mass in the hedgehog configuration in the two phase chiral bag model.
We discuss the chiral properties, such as the fractional baryon number and the chiral Casimir energy, by using the Debye expansion for the analytical calculation and the Strutinsky's smearing method for the numerical computation.
It is shown that the fractional baryon number carried by massive quarks in the vacuum is canceled by that in the meson sector.
A finite term of the chiral Casimir energy is obtained with subtraction of the logarithmic divergence term. 
\end{abstract}

\section{Introduction}

Strangelets of finite volume quark matter with strangeness has been one of the most interesting subjects of exotic particles in  hadron and quark  physics \cite{Bodmer, Witten, Farhi, Weber}.
In recent studies of strangelets, a model has been presented for a discussion of chiral symmetry breaking inside a quark droplet of finite volume quark matter \cite{Kiriyama_Hosaka, Yasui, Yasui2, Yasui3}.
There, in addition to quark confinement for a finite volume system, it was allowed that the quarks acquire a finite mass by dynamical chiral symmetry breaking.
In this picture, it was assumed that the quark interaction was provided by the Nambu--Jona-Lasinio (NJL) type interaction inside a quark droplet.
In earlier works, the quark wave function was given by the MIT bag in Refs.~\cite{Kiriyama_Hosaka, Yasui, Yasui2}.
Later, in order to overcome the chiral symmetry breaking at the bag surface, the chiral bag model was introduced in Ref.~\cite{Yasui3}.
The latter model was called as the NJL chiral bag model.
Such an idea was first presented by T.~Kunihiro in 1983 \cite{Kunihiro1, Kunihiro2}.

In the discussion of strangelets by the NJL chiral bag model, the hedgehog ansatz was used for the pion and $ud$ quarks to include non-linear interaction of pions and quarks \cite{Yasui3}.
In the history of the study of the chiral bag model since \cite{Inoue, Chodos}, effects of vacuum polarization, such as the anomalous baryon number and the chiral Casimir effects, have been discussed extensively for massless quarks \cite{Brown_Rho, Brown_Rho_Vento, Vento, Thomas, Goldstone, Mulders}.
It was shown that the correct behavior of the baryon number was provided by the contribution from the sea quarks  \cite{Goldstone, Zahed84, Zahed}.
In the development of the chiral bag model, several techniques for calculation of the sea quark contribution was also developed.
In Ref. \cite{Goldstone}, the fractional baryon number was derived by using the Multiple Reflection Expansion method.
After that, the Debye expansion was used \cite{Zahed84, Zahed}.
The Casimir energy is also affected by the pion cloud.
It was shown that the chiral Casimir energy had a term of logarithmic divergence \cite{Vepstas, Hosaka_Toki}.
However, these properties for massive quarks have not been investigated so far.
In the framework of the NJL chiral bag model, the dynamical quark mass is an order parameter for the chiral symmetry breaking in the chiral bag.
In order to search the real stable state in the energy variation with respect to the dynamical quark mass, we need to clarify the effect of vacuum polarization for the finite quark mass in the hedgehog configuration in the chiral bag.
This is the motivation in the present paper.

We organize this paper as follows.
In Section 2, we propose a model lagrangian of the chiral bag with massive quarks.
In this paper, we describe the quarks with a finite Dirac mass, which is generated through the NJL type interaction in the chiral bag.
In Section 3, we discuss a fractional baryon number for massive quarks by using both of numerical computation and analytical formulation.
In Section 4, we discuss the chiral Casimir energy.
Deriving the terms of logarithmic divergence 
the finite term of the chiral Casimir energy is obtained numerically.
We summarize our discussion in Section 5.

\section{The chiral bag with massive quarks}

In the NJL chiral bag model \cite{Yasui3, Kunihiro1, Kunihiro2}, the quark mass in a finite volume system is induced by the NJL point-like interaction \cite{Nambu}.
In the NJL chiral bag model, we consider the $ud$ quark sector with a finite quark mass induced by the mean field approximation in the scalar channel.
The lagrangian in the $ud$ quark sector is written as
\begin{eqnarray}
 {\cal L} &=& 
                         \bar{\psi}( i \partial\hspace{-0.2cm}/ - \hat{m})\psi
                   \theta(R-r)
                  -\frac{1}{2}\bar{\psi} U^{\gamma_{5}} \psi \delta(r-R),
 \label{eq : NJL_bag}
\end{eqnarray}
which includes a boundary condition explicitly. 
Here, $\psi=(u,d)^{t}$ is the $ud$ quark fields, and the $ud$ quark mass matrix is given by $ \hat{m} = \mbox{diag}(m_{u},m_{d}) $ with  flavor symmetry $ m = m_{u} = m_{d}$.
The $ud$ quark mass is the sum of the current mass and the constituent mass which is generated in the non-zero expectation value of the quark scalar condensate $\bar{q}q$.
In general, the constituent quark mass can be position dependent.
Here, in order to simplify the essential discussion of the finite quark mass, we treat the quark mass as a constant.
For a bag configuration, we assume a static spherical bag with radius $R$.
The step function is multiplied in the first term in order to confine quarks inside the bag.
Here, $r$ is a distance from the center of the bag.
The second term with the delta function realizes the chiral invariant boundary condition
 at the bag surface, where we define
\begin{eqnarray}
  U^{\gamma_{5}} &=& e^{i\vec{\tau}\cdot\vec{\pi} \gamma_{5}},
\end{eqnarray}
with the $\pi$ meson field $\vec{\pi}$ and the Pauli matrix $\vec{\tau}$.
In this paper, we do not explicitly show the lagrangian in the pion sector, since our current interest is to study the chiral vacuum polarization of the quarks in the bag.

In order to consider the non-linear effect of the pion, we assume the hedgehog ansatz in the $\pi$ meson sector, where the pion field conserves the grand spin $\vec{K} = \vec{J} + \vec{I}$ with total angular momentum $\vec{J}$ and isospin $\vec{I}$.
The hedgehog pion field for $r>R$ is written as
\begin{eqnarray}
  \vec{\pi} = F(r) \vec{n},
\end{eqnarray}
with a chiral angle $F(r)$ and a unit radial vector $\vec{n}$ in the real space \cite{Skyrme, Adkins}.

According to the $\pi$ meson sector, we introduce the hedgehog basis set in the $ud$ quark sector \cite{Mulders, Hosaka}.
We construct the quark wave function $\psi^{(\kappa)}$ for natural ($\kappa=+$) and unnatural ($\kappa=-$) assignment for grand spin $K$, respectively;
\begin{eqnarray}
  \psi^{(+)} = 
\left(
\begin{array}{c}
 a_{0} j_{K}(pr) | 0 \rangle + a_{1} j_{K}(pr) | 1 \rangle \\
 a_{2} j_{K+1}(pr) | 2 \rangle + a_{3} j_{K-1}(pr) | 3 \rangle
\end{array}
\right),
\end{eqnarray}
and
\begin{eqnarray}
  \psi^{(-)} = 
\left(
\begin{array}{c}
 b_{2} j_{K+1}(pr) | 2 \rangle + b_{3} j_{K-1}(pr) | 3 \rangle \\
 b_{0} j_{K}(pr) | 0 \rangle + b_{1} j_{K}(pr) | 1 \rangle
\end{array}
\right).
\end{eqnarray}
The coefficients $a_{i}$ and $b_{i}$ ($i=0,\cdots,3$) are determined by satisfying the equation of motion of the $ud$ quark for $\kappa=\pm$, respectively.
$j_{K}(pr)$ is the spherical Bessel function and $p$ is the $ud$ quark momentum.
Here, the two-component spinors $ | 0 \rangle, \cdots,  | 3 \rangle $ are given by
\begin{eqnarray}
  | 0 \rangle &=& Y_{KM}(\theta, \phi) \chi_{0}^{0},
 \nonumber \\
  | 1 \rangle &=& \sum_{\mu=-1,0, 1} ( K M-\mu 1 \mu | K M ) Y_{K M-\mu}(\theta, \phi) \chi_{\mu}^{1},
 \nonumber \\
  | 2 \rangle &=& \sum_{\mu=-1,0, 1} ( K+1 M-\mu 1 \mu | K M ) Y_{K+1 M-\mu}(\theta, \phi) \chi_{\mu}^{1},
 \nonumber \\
  | 3 \rangle &=& \sum_{\mu=-1,0, 1} ( K-1 M-\mu 1 \mu | K M ) Y_{K-1 M-\mu}(\theta, \phi) \chi_{\mu}^{1},
\end{eqnarray}
where $Y_{L \, M}(\theta, \phi)$ is the spherical harmonics with spherical coordinate $(\theta, \phi)$.
$\chi_{\mu}^{G}$ are eigenstates of $\vec{G} = \vec{S} + \vec{I} = \vec{\sigma}/2+\vec{\tau}/2$,
\begin{eqnarray}
 && \chi_{0}^{0} = \frac{1}{\sqrt{2}}( | \uparrow \rangle  | d \rangle - | \downarrow \rangle  | u \rangle ),
 \nonumber \\
 && \chi_{1}^{1} = | \uparrow \rangle  | u \rangle,
 \nonumber \\
&&  \chi_{0}^{1} = \frac{1}{\sqrt{2}}( | \uparrow \rangle  | d \rangle + | \downarrow \rangle  | u \rangle ),
 \nonumber \\
&&  \chi_{-1}^{1} = | \downarrow \rangle  | d \rangle.
\end{eqnarray}
The sign of naturalness $\kappa$ corresponds to the parity $P=(-)^{K+\kappa}$.

In order to satisfy the equation of motion, the coefficients $a_{i}$ for naturalness assignment $\kappa=+$ are subjected to
\begin{eqnarray}
\left(
\begin{array}{cccc}
 E-m &  0      & -iPp  & -iQp  \\
 0     &  E-m  & -iQp  & iPp  \\
 iPp  &  iQp  & E+m   &  0 \\
  iQp & -iPp & 0        & E+m
\end{array}
\right)
\left(
\begin{array}{c}
 a_{0} \\
 a_{1} \\
 a_{2} \\
 a_{3}
\end{array}
\right) = 0,
  \label{eq : matrix_equation}
\end{eqnarray}
where we define
\begin{eqnarray}
&& P = \sqrt{\frac{K+1}{2K+1}},
 \nonumber \\
&& Q = \sqrt{\frac{K}{2K+1}},
\\ \nonumber
&& E = \sqrt{ p^{2} + m^{2} }.
\end{eqnarray}
Then, we obtain two independent solutions
\begin{eqnarray}
\vec{a'} &=&
\left(
\begin{array}{c}
 a'_{0} \\
 a'_{1} \\
 a'_{2} \\
 a'_{3}
\end{array}
\right) = N'
\left(
\begin{array}{c}
 \frac{p}{E-m} iP \\
 \frac{p}{E-m} iQ \\
 1 \\
 0
\end{array}
\right),
 \nonumber \\
\vec{a''} &=&
\left(
\begin{array}{c}
 a''_{0} \\
 a''_{1} \\
 a''_{2} \\
 a''_{3}
\end{array}
\right) = N''
\left(
\begin{array}{c}
 \frac{p}{E-m} iQ \\
 -\frac{p}{E-m} iP \\
 0 \\
 1
\end{array}
\right). 
\end{eqnarray}
Here, the normalization constants $N'$ and $N''$ are determined by the condition $\int d^{3}x \psi^{(+)\dag} \psi^{(+)}=1$,
\begin{eqnarray}
 N'^{-2} &=& R^{3} \frac{E}{E-m} 
                   \left[
				            j_{K}(pR)^{2} + j_{K+1}(pR)^{2}
					     -\frac{1}{pR}\left( 2(K+1) - \frac{m}{E} \right) j_{K}(pR)j_{K+1}(pR)
				   \right],
 \nonumber \\
N''^{-2} &=& R^{3} \frac{E}{E-m} 
                   \left[
				            j_{K-1}(pR)^{2} + j_{K}(pR)^{2}
					     -\frac{1}{pR}\left( 2K + \frac{m}{E} \right) j_{K-1}(pR)j_{K}(pR)
				   \right].
\end{eqnarray}
The final solution is expressed as a linear combination of $\vec{a'}$ and $\vec{a''}$,
\begin{eqnarray}
 \vec{a} =c'\vec{a'} + c''\vec{a''} =
 c'N'
 \left(
\begin{array}{c}
 \frac{p}{E-m} iP \\
 \frac{p}{E-m} iQ \\
 1 \\
 0
\end{array}
\right)  +
c''N''
\left(
\begin{array}{c}
 \frac{p}{E-m} iQ \\
 -\frac{p}{E-m} iP \\
 0 \\
 1
\end{array}
\right),
\label{eq : solution+}
\end{eqnarray}
with constants $c'$ and $c''$.
We obtain a solution for the unnaturalness assignment $\kappa=-$ in the same way.

The eigenenergy of the hedgehog $ud$ quark is given by the boundary condition at the bag surface $r=R$,
\begin{eqnarray}
  i\vec{n}\cdot\vec{\gamma} \psi^{(\kappa)} = - e^{i\vec{\tau}\cdot\vec{n} F(R)\gamma_{5}} \psi^{(\kappa)}.
  \label{eq : boundary_vector_1}
\end{eqnarray}
By substituting the solution of Eq.~(\ref{eq : solution+}), we obtain an equation for the eigenvalue, which is given by
\begin{eqnarray}
   && \cos F(R) \left\{ j_{K}(pR)^{2} - \left(\frac{E-\kappa m}{p}\right)^{2} j_{K+1}(pR) j_{K-1}(pR) \right\}
   \nonumber \\
 &&- \kappa \frac{E-\kappa m}{p} j_{K}(pR) \left\{ j_{K+1}(pR) - j_{K-1}(pR) \right\}
   \nonumber \\
 &&+ \frac{E-\kappa m}{p} \frac{\sin F(R)}{2K+1} j_{K}(pR) \left\{ j_{K+1}(pR) + j_{K-1}(pR) \right\} = 0.
  \label{eq : boundary_vector_2}
\end{eqnarray}
The equation for $K=0$ is obtained by setting $j_{-1}(pR)=0$.
Eq.~(\ref{eq : boundary_vector_2}) determines the energy spectrum of a quark in the bag.
We mention that there is a symmetry of the energy level in the positive and negative energy
\begin{eqnarray}
   E_{K^{P}}(F) = -E_{K^{-P}}(-F).
\end{eqnarray}
This relation has its origin in the invariance of the lagrangian (\ref{eq : NJL_bag}) under the transformation $U \rightarrow U^{\ast}$ or $F \rightarrow -F$.
Another symmetry under $F \rightarrow F+\pi$ and $\kappa \rightarrow -\kappa$ is conserved in the massless case.
However, this symmetry is not conserved for the finite mass.
This is due to an asymmetry of the quark energy levels, which is shown in Fig.~1 in \cite{Yasui3}.

\section{Baryon number conservation}

In the limit of the zero bag radius, the meson field in the Skyrme lagrangian gives a mapping from $R^{3} \sim S^{3}$ to $SU(2) \sim S^{3}$ by imposing a boundary condition $U=1$ at $r \rightarrow \infty.$
The winding number in the mapping is identified with the baryon number \cite{Skyrme, Adkins}.
The fractional baryon number due to the pion cloud is given as \cite{Skyrme, Adkins}
\begin{eqnarray}
   B_{\pi} = - \frac{1}{\pi} \left[ F - \frac{1}{2} \sin 2F \right],
   \label{eq : meson_baryon_number}
\end{eqnarray}
where $F$ is the chiral angle at the bag surface.
In the pure Skyrmion, the chiral angle is given as $n\pi$ with integer $n$.

On the other hand, in the chiral bag with finite bag radius, the chiral angle is not generally equal to $n\pi$.
Therefore, the meson carries only a fractional baryon number.
There, it has been known that the total baryon number is composed of the fractional baryon numbers of the pion, of the vacuum and of the valence quarks in a bag \cite{Mulders, Hosaka}.
The baryon number carried by the vacuum quarks is defined as
\begin{eqnarray}
   B_{q}(m, F) &=& \frac{1}{2} \langle {\cal O} | [\psi^{\dag}, \psi] |\cal{O}\rangle
  \nonumber \\ 
   &=& - \frac{1}{2} \lim_{t \rightarrow 0} \sum_{n} \mbox{sgn}(E_{n}) e^{-t|E_{n}|}.
    \label{eq : baryon_number0}
\end{eqnarray}
Here, $|\cal{O}\rangle$ is a vacuum state of an empty bag filled with negative energy quarks, $E_{n} = E_{n}(m, F)$ an $n$-th state energy of a quark with mass $m$, and at the chiral angle $F$ at the bag surface.
The quantum number $n$ labels the grand spin and parity.
The sum is taken over for all the quark states $n$ with positive and negative energies in the bag.
In order to obtain the convergence in the sum, an exponential type regularization is multiplied.

In the studies of the chiral bag model with massless quarks, it was shown that the baryon number of the vacuum quarks was canceled by that of the meson cloud \cite{Brown_Rho, Brown_Rho_Vento, Vento, Goldstone, Mulders, Hosaka_Toki, Zahed}.
Accordingly, the total baryon number is a conserved quantity.
This result should also be the case for the chiral bag with massive quarks.
It is expected that the baryon number conservation is not affected by the finite quark mass, because $U(1)_{B}$ symmetry is conserved in our lagrangian (\ref{eq : NJL_bag}).
In the following, we show explicitly the cancellation of the fractional baryon number between the vacuum massive quarks and the pion cloud.
We use both numerical computation and analytical procedure.

First, we show the numerical computation.
In the studies of chiral bag model with massless quarks, the regularization, such as the Gaussian type \cite{Vepstas}, heat kernel type \cite{Hosaka_Toki}, have been used with success.
In this paper, we use the Strutinsky's smearing method \cite{Wust, Vepstas_Jackson}.
This method has an advantage that the states necessary for computation are limited up to the grand spin $K_{max} \sim 40$ in order to obtain a good convergence, while we need $K_{max} \sim 100$ at least  for the other regulators.
We rewrite the baryon number (\ref{eq : baryon_number0}) by introducing a delta function for a density of states  to pick up the discrete levels
\begin{eqnarray}
     B_{q}(m, F) = - \frac{1}{2} \int_{-\infty}^{\infty} dx \sum_{n} \mbox{sgn}(E_{n}) \delta(x-E_{n}R)
\end{eqnarray}
Then, we replace the delta function by a gaussian function,
\begin{eqnarray}
     B_{q}(m, F) = - \frac{1}{2} \int_{-\infty}^{\infty} dx \rho_{x}(m, F),
\end{eqnarray}
where we define a function for a density of states
\begin{eqnarray}
  \rho_{x}(m, F) = \sum_{n} \mbox{sign}(E_{n})
                                                     \frac{ e^{ -(x-E_{n}R)^{2}/\gamma^{2} } }{\gamma \sqrt{\pi}},
 \label{eq : chiral_state_density}
\end{eqnarray}
where $\gamma \simeq 2-3$ is a smearing parameter.
We obtain $B_{q}=0$ at $F=0$, since the energy spectrum is symmetric for positive and negative energies.
In oder to obtain a rapid convergence in the sum, it is convenient to define the difference of $\rho_{x}(m, F)$ between finite $F$ and $F=0$,
\begin{eqnarray}
 \tilde{\rho}_{x}(m, F) =  \rho_{x}(m, F) - \rho_{x}(m, 0).
 \label{eq : chiral_state_density2}
\end{eqnarray}
Therefore, the baryon number defined by Eq.~(\ref{eq : baryon_number0}) is given by an alternative formulation
\begin{eqnarray}
   B_{q}(m, F) = \int_{-\infty}^{\infty} dx \tilde{\rho}_{x}(m, F).
   \label{eq : baryon_number1}
\end{eqnarray}
For a numerical computation, we restrict the range of the $x$-integral in a finite interval $x \in [-x_{max}, x_{max}]$.
\begin{eqnarray}
   B_{q}(m, F) = \int_{-x_{max}}^{x_{max}} dx \tilde{\rho}_{x}(m, F).
\end{eqnarray}
We use $x_{max} \simeq 20$ to obtain sufficient convergence.
Carrying out numerical computation, we obtain the result which is given by
\begin{eqnarray}
 B_{q}(m, F) = \frac{1}{\pi} \left[F-\frac{1}{2} \sin 2F \right]
 \label{eq : baryon_number2}
\end{eqnarray}
for $F \in [F_{0}, F_{0}+\pi]$.
The value of $F_{0}$ depends on the quark mass; $F_{0}/\pi=-0.5$, $-0.695$, $-0.816$ and $-0.878$ for $mR=0$, $1$, $2$ and $3$, respectively.
The equation (\ref{eq : baryon_number2}) is independent of quark mass $m$,
 and coincides with the opposite sign of Eq.~(\ref{eq : meson_baryon_number}) in the meson sector.
Therefore, the fractional baryon number by meson and vacuum quark cancel each other for any quark mass.

The total baryon number is equal to $A$ for a quark droplet with $3A$ valence quarks for any chiral angle.
Let us investigate the case of a baryon.
The baryon number carried by valence and vacuum quarks for chiral angle $F_{0} < F < 0 $ is given as a sum of the valence quark and the fractional  baryon number
\begin{eqnarray}
  B(m,F)_{val+sea} = 1+\frac{1}{\pi} \left[F-\frac{1}{2} \sin 2F \right].
\end{eqnarray}
For $-\pi < F < F_{0} $, the valence quark is absent since the $0^{+}$ state is absorbed in the vacuum.
Then, the baryon number is obtained by substituting $F + \pi$ instead of $F$ from a periodicity, 
\begin{eqnarray}
  B(m,F)_{val+sea} &=& \frac{1}{\pi} \left[(F+\pi)-\frac{1}{2} \sin 2(F+\pi) \right]
   \nonumber \\
                              &=& 1+\frac{1}{\pi} \left[F-\frac{1}{2} \sin 2F \right].
\end{eqnarray}
Therefore, the total baryon number is one for any chiral angle.
In the same way, the conservation of baryon number holds for any quark droplets.

Second, we present an analytical calculation of the baryon number (\ref{eq : baryon_number0}) by using the Debye expansion.
This method was originally developed in the chiral bag with massless quark \cite{Zahed84, Zahed}.
We apply this formalism to the massive case.
We rewrite Eq.~(\ref{eq : baryon_number0}) in terms of the quark propagator as in \cite{Zahed84, Zahed}.
Then, we obtain
\begin{eqnarray}
  B_{q}(m, F) = \frac{1}{2} \lim_{\tau \rightarrow 0_{+}} \int_{-\infty}^{\infty} \frac{dx}{2i\pi} e^{\-i\tau x} \sum_{K=0}^{\infty} (2K+1) \frac{d}{dx} \ln \left[ \frac{S_{K}(m, F;ix)}{S_{K}^{\ast}(m, F;ix)} \right]
  \label{eq : fractional_baryon1}
\end{eqnarray}
Here, $S_{K}(m, F;p)$ is given as a product of the left hand side of the eigenvalue equation Eq.~(\ref{eq : boundary_vector_2}) for natural ($\kappa=+1$) and unnatural ($\kappa=-1$) states.
For the sake of using the Debye expansion, we substitute an imaginary number $ix$ for $pR$ in the spherical Bessel function.
We explicitly show
\begin{eqnarray}
  S_{0}(F;ix) = \cos F \left( j_{1}(ix)^{2} - j_{0}(ix)^{2} \right) - 2 \frac{E \sin F + \bar{m}}{p} j_{0}(ix) j_{1} (ix)
  \label{eq : S_equation_0}
\end{eqnarray}
for $K=0$, and
\begin{eqnarray}
 S_{K}(m, F;ix) &=& \left[ \frac{\bar{E}}{\bar{p}} \cos F \left( j_{K}(ix)^{2} - j_{K+1}(ix)j_{K-1}(ix) \right)
                             + \frac{\sin F}{2K+1} j_{K}(ix) \left( j_{K+1}(ix)+j_{K-1}(ix) \right) \right]^{2}
\label{eq : S_equation_1} \nonumber \\
   &&+ \left[ \frac{\bar{m}}{\bar{p}} \left( j_{K}(ix)^{2} + j_{K+1}(ix)j_{K-1}(ix) \right) - j_{K} \left( j_{K+1}(ix) - j_{K-1}(ix) \right)  \right]^{2}.
\end{eqnarray}
for $K\neq0$.
For a short notation, we write $\bar{m}=mR$, $\bar{p}=x$ and $\bar{E}=\sqrt{\bar{m}^{2}+x^{2}}$.
The $x$-integral picks up an eigenvalue as a residue instead of solving directly the boundary condition.
We note that Eq.~(\ref{eq : fractional_baryon1}) is an exact formulation, but it is not practical to consider the $x$-integral and the sum over $K$ without approximation.
In order to obtain the final result, it is sufficient to use an asymptotic behavior of the quark energy spectrum.

In the following manipulation, we discuss the $K=0$ and $K \ge 1$ components separately for convenience. 
First, we consider the $K=0$ component, which is defined by
\begin{eqnarray}
  B_{q}^{(K=0)}(m, F) = \frac{1}{2} \lim_{\tau \rightarrow 0_{+}} \int_{-\infty}^{\infty} \frac{dx}{2i\pi} e^{\-i\tau x}  \frac{d}{dx} \ln \left[ \frac{S_{0}(m, F;ix)}{S_{0}^{\ast}(m, F;ix)} \right].
  \label{eq : baryon_number_K0}
\end{eqnarray}
It is convenient to write the function $S_{0}(m, F; ix)$ in a polar coordinate
\begin{eqnarray}
 S_{0}(m, F;ix) = R_{0}(x) e^{i \Phi_{0}(x)}.
\end{eqnarray}
By using the asymptotic form
\begin{eqnarray}
  j_{1}(ix) &\simeq& \frac{i}{2x} \left( e^{x} + e^{-x} \right),
  \label{eq : asymptotic_Bessel}
\end{eqnarray}
we obtain
\begin{eqnarray}
  \Phi_{0}(x) \simeq \arctan 
             \left[ \frac{e^{2x}-e^{-2x}}{e^{2x}+2^{-2x}} \frac{\sqrt{x^{2}+\bar{m}^{2}} \sin F + \bar{m}}{x\cos F} \right].
\end{eqnarray}
Then, it is straightforward to perform the integral, giving
\begin{eqnarray}
  B_{q}^{(K=0)}(m, F) \simeq \frac{F}{\pi},
  \label{eq : baryon_number_1} 
\end{eqnarray}
for any quark mass $m$.
 
Second, for $K \ge 1$, we use the Debye expansion for the modified Bessel function in estimation of Eq.~(\ref{eq : S_equation_1}).
In our notation, the modified Bessel function $I_{\nu}(x)$ is defined by
\begin{eqnarray}
   j_{K}(ix) = \sqrt{\frac{\pi}{2x}} i^{K} I_{\nu}(x),
\end{eqnarray}
with $\nu=K+1/2$.
The Debye expansion is a uniform asymptotic expansion
  with no constraint between $\nu$ and $x$.
The Debye expansion gives
\begin{eqnarray}
  I_{\nu}(x) \simeq \frac{\Gamma(1/2)}{\pi \sqrt{2t}} e^{f_{\nu}(x)},
  \label{eq : Debye}
\end{eqnarray}
in the lowest order,
where we define $t=\sqrt{\nu^{2}+t^{2}}$ and $f_{\nu}(x)=t-\nu \, \mbox{sinh}^{-1} (\nu/x)$.

Now, let us define
\begin{eqnarray}
  B_{q}^{(K \ge 1)}(m, F) = \frac{1}{2} \lim_{\tau \rightarrow 0_{+}} \int_{-\infty}^{\infty} \frac{dx}{2i\pi} e^{\-i\tau x}  \sum_{K\ge 1} \frac{d}{dx} \ln \left[ \frac{S_{K}(m, F;ix)}{S_{K}^{\ast}(m, F;ix)} \right].
  \label{eq : baryon_number_K>0}
\end{eqnarray}
Here, we write $S_{K}(m, F; ix)$ in a polar coordinate
\begin{eqnarray}
 S_{K}(m, F; ix) = R_{K}(x) e^{i \Phi_{K}(x)}.
\end{eqnarray}
Then, we apply the Debye expansion (\ref{eq : Debye}).
After a little tedious calculation, we pick up only the terms of the leading order of  $\bar{m}$, and obtain
\begin{eqnarray}
 \Phi_{K}(x) \simeq \arctan \left[ -\frac{1}{2K+1} \left( \frac{\nu x}{2t^{3}} \sin 2F - \frac{4\nu \bar{m}}{t} \right) \right].
\end{eqnarray}
We consider that the quark mass $m$ is smaller as compared with the momentum $p$.
Indeed, this is a good approximation, since we are interested in the asymptotic behavior of the quark energy spectrum. 
The $x$-integral for $\Phi_{K}(x)$ is integrated out and the integral is determined only by $\Phi_{K}(\infty) - \Phi_{K}(-\infty)$.
Concerning the term which has no quark mass, we use an identity
\begin{eqnarray}
   \sum_{K\ge1} \frac{\cos( \tau \nu z )}{\nu} = -\ln \tan (\tau z/4) -2\cos(\tau z/2),
   \label{eq : sum}
\end{eqnarray}
and
\begin{eqnarray}
  \lim_{\tau \rightarrow 0_{+}} \int_{-\infty}^{\infty} \frac{dx}{2\pi i} e^{-i\tau x} \sum_{K \ge1} \frac{d}{dx} \left[ -i \frac{\nu x}{2t^{3}} \right] &=& -\frac{1}{2\pi}.
\end{eqnarray}
Concerning the term proportional to $\bar{m}$, we use
\begin{eqnarray}
  \lim_{\tau \rightarrow 0_{+}} \int_{-\infty}^{\infty} \frac{dx}{2\pi i} e^{-i\tau x} \sum_{K \ge1} \frac{d}{dx} \frac{4\nu \bar{m}}{t} &=& 0,
\end{eqnarray}
which indicates the mass term does not contribute to the integral.
Consequently, we obtain the result
\begin{eqnarray}
  B_{q}^{(K \ge 1)}(m, F) = - \frac{\sin 2F}{2\pi}.
  \label{eq : baryon_number_2}
\end{eqnarray}
Finally, by Eqs.~(\ref{eq : baryon_number_1}) and (\ref{eq : baryon_number_2}), the fractional baryon number obtained in the analytical procedure coincides with the numerical result Eq.~(\ref{eq : baryon_number2}).

\section{Chiral Casimir energy}

The chiral Casimir energy arises as a result of the modification of quark energy levels for finite pion cloud.
Following the regularization scheme in the baryon number, the chiral Casimir energy is defined by
\begin{eqnarray}
  E_{C}(m, F) &=& \frac{1}{2} \langle {\cal O} | [\psi^{\dag}, H \psi] | {\cal O} \rangle
   \label{eq : chiral_Casimir_energy}
  \nonumber \\
   &=&- \frac{1}{2} \lim_{t \rightarrow 0_{+}} \sum_{n} \mbox{sign}(E_{n}) E_{n}  e^{-t|E_{n}|},
\end{eqnarray}
as in Refs.~\cite{Brown_Rho, Brown_Rho_Vento, Vento, Mulders, Hosaka_Toki_96, Hosaka_Toki, Hosaka}.

We show the numerical procedure by using the Strutinsky's smearing method.
By using the Gaussian function (\ref{eq : chiral_state_density}), we rewrite Eq.~(\ref{eq : chiral_Casimir_energy}) as
\begin{eqnarray}
  E_{C}(m, F) = \int_{-\infty}^{\infty} dx x \rho_{x}(m, F).
\end{eqnarray}
By choosing the reference point at $F=0$, 
we obtain an alternative formulation
\begin{eqnarray}
  \Delta E_{C}(m, F) &=& E_{C}(m, F) - E_{C}(m, 0)
 \nonumber \\
 &=& \int_{-\infty}^{\infty} dx x \tilde{\rho}_{x}(m, F).
\end{eqnarray}
Restricting the range of integral in a finite range of $[-x_{max}, x_{max}]$ with sufficient convergence, 
 we obtain,
 \begin{eqnarray}
  \Delta E_{C}(m, F) = \int_{-x_{max}}^{x_{max}} dx x \tilde{\rho}_{x}(m, F).
  \label{eq : chiral_Casimir_energy2}
\end{eqnarray}
In order to achieve a rapid convergence of the integral, it is much more practical to use the partial integration\footnote{We use a relationship
$
   \int_{-\infty}^{\infty} dx x \tilde{\rho}_{x}(m, F)
   = - \int_{-\infty}^{\infty} dx \frac{x^{2}}{2} \tilde{\rho}'_{x}(m, F)
$
with $\tilde{\rho}_{x}(m,F) \rightarrow 0$ for $|x|\rightarrow \pm \infty$.
}.
We mention that we need $x_{max} \sim 40$ to obtain the convergence in the calculation.
More states are necessary for the chiral Casimir energy as compared for the baryon number ($x_{max} \sim 20$).

For massless quarks, it has been known that there is a logarithmic divergence term in the Casimir energy (\ref{eq : chiral_Casimir_energy}).
It was shown that the divergence term is proportional to $\sin^{2} F$ \cite{Brown_Rho, Brown_Rho_Vento, Vento, Zahed84, Zahed, Hosaka_Toki_96,  Hosaka_Toki, Hosaka}.
Here, we also derive the logarithmic divergence term for massive quark by using the Debye expansion developed in \cite{Zahed84, Zahed}.
Following the procedure in \cite{Zahed84, Zahed}, we rewrite the Casimir energy as 
\begin{eqnarray}
 \Delta E_{C}(m,F) = \frac{1}{2R} \lim_{\tau \rightarrow 0_{+}} \int_{-\infty}^{\infty}
 \frac{dx}{2\pi} x e^{i\tau x} \sum_{K=0}^{\infty} (2K+1) \frac{d}{dx} \ln \left| \frac{S_{K}(m, F;ix)}{S_{K}(m, 0;ix)} \right|.
\end{eqnarray}
Just as in the case of the baryon number, we consider the calculation for $K=0$ and $K \ge 1$, separately for convenience.

First, we investigate the $K=0$ sector,
\begin{eqnarray}
 \Delta E_{C}^{(K=0)}(m,F) = \frac{1}{2R} \lim_{\tau \rightarrow 0_{+}} \int_{-\infty}^{\infty}
 \frac{dx}{2\pi} x e^{i\tau x} \frac{d}{dx} \ln \left| \frac{S_{0}(m, F;ix)}{S_{0}(m, 0;ix)} \right|.
\end{eqnarray}
We apply the asymptotic form of the spherical Bessel function (\ref{eq : asymptotic_Bessel}).
By expanding $ \left| S_{0}(m, F;ix)/S_{0}(m, 0;ix) \right|$ in the lowest order of $m$, we obtain
\begin{eqnarray}
 \Delta E_{C}^{(K=0)}(m,F) \simeq \frac{1}{2R} \lim_{\tau \rightarrow 0_{+}} \int_{-\infty}^{\infty}
 \frac{dx}{2\pi} x e^{i\tau x} \frac{d}{dx} \left[ \cos^{2}F + \sin F \tanh^{2}(2x) + \sin F \, \frac{2\bar{m}}{x} \tanh^{2}(2x) \right].
\end{eqnarray}
The first two terms in the integrand, which does not contain quark mass, are given in \cite{Zahed84, Zahed}. 
The integral of the third term proportional to $\bar{m}$ is easily shown to be equal to zero.
Consequently, we obtain the result
\begin{eqnarray}
 \Delta E_{C}^{(K=0)}(m,F) \simeq \frac{F^{2}}{4\pi R}. 
\end{eqnarray}
The divergence term does not appear in the $K=0$ sector.

Second, we investigate the sum over $K \ge 1$,
\begin{eqnarray}
 \Delta E_{C}^{(K \ge 1)}(m, F) = \frac{1}{2R} \lim_{\tau \rightarrow 0_{+}} \int_{-\infty}^{\infty}
 \frac{dx}{2\pi} x e^{i\tau x} \sum_{K=1}^{\infty} (2K+1) \frac{d}{dx} \ln \left| \frac{S_{K}(m, F;ix)}{S_{K}(m, 0;ix)} \right|.
\end{eqnarray}
After a little troublesome calculation by using the Debye expansion and expanding in the lowest order of $\bar{m}$, we obtain
\begin{eqnarray}
  \ln \left| \frac{S_{K}(m, F;ix)}{S_{K}(m, 0;ix)} \right|
   &\simeq& -\sin^{2} F \left[ \frac{(1-L_{+}L_{-})^{2}}{(1+L_{+}^{2})(1+L_{-}^{2})} + \frac{1}{(2K+1)^{2}} \frac{(L_{+}-L_{-})^{2}}{(1+L_{+}^{2})(1+L_{-}^{2})} \right]
    \nonumber \\
   &&+ \frac{\bar{m}^{2}}{2x^{2}} \sin^{2}F \frac{(1-L_{+}L_{-})^{2}}{(1+L_{+})^{2}(1+L_{-})^{2}}
  \nonumber \\
    &=& -\frac{1}{2} \sin^{2}F \frac{1}{t^{2}} \left[ \left(2-\frac{\nu^{2}}{t^{2}}\right) + \left(1-\frac{\nu^{2}}{t^{2}} \right)\left(2-\frac{3\nu^{2}}{t^{2}}\right)\frac{1}{t} + {\cal O}(1/t^{2}) \right]
  \nonumber \\
   &&+ \frac{\bar{m}^{2}}{2} \sin^{2}F \frac{1}{4} \left( 1-\frac{\nu^{2}}{t^{2}} \right)
             \frac{1}{t^{2}} \left[ 1+\left( 1-\frac{3\nu^{2}}{t^{2}}\frac{1}{t} \right) + {\cal O}(1/t^{2})  \right].
\end{eqnarray}
Here, we define $L_{\pm}=I_{\nu \pm 1}/I_{\nu}$ for a short notation. 
The first term, which does not contain quark mass, coincides with the case of massless quarks \cite{Zahed84, Zahed}.
The second term proportional to $\bar{m}$ is further analyzed as followings.
By taking a derivative with respect to $x$, we obtain
\begin{eqnarray}
  x \frac{d}{dx} \ln \left| \frac{S_{K}(m, F;ix)}{S_{K}(m, 0;ix)} \right|
    &\simeq& -\frac{1}{2} \sin^{2}F \left( -4\frac{x^{4}}{t^{6}} - \frac{6}{t^{3}} + \frac{31\nu^{2}}{t^{5}} - \frac{46\nu^{4}}{t^{7}} + \frac{21\nu^{6}}{t^{9}} \right)
 \nonumber \\
   &&+ \frac{\bar{m}^{2}}{8} \sin^{2}F \left( -\frac{2}{t^{2}} - \frac{3}{t^{3}} + \frac{6\nu^{2}}{t^{4}} + \frac{23\nu^{2}}{t^{5}} - \frac{4\nu^{4}}{t^{6}} - \frac{41\nu^{4}}{t^{7}} + \frac{21\nu^{6}}{t^{9}} \right).
\end{eqnarray}
Then, introducing new variables $h=t/\nu$ and $z=x/\nu$, we have
\begin{eqnarray}
   \Delta E_{C}^{(K \ge 1)}(m, F) 
   &\simeq&
   \frac{\sin^{2}F}{\pi R} \lim_{\tau \rightarrow 0_{+}} \int_{0}^{\infty} dz \frac{2z^{4}}{h^{6}} \sum_{K=1}^{\infty} \cos( \tau \nu z )
 \nonumber \\
   &&+ \frac{\sin^{2}F}{\pi R} \lim_{\tau \rightarrow 0_{+}} \int_{0}^{\infty} dz \left( \frac{6}{h^{3}} - \frac{31}{h^{5}} + \frac{46}{h^{7}} - \frac{21}{h^{9}} \right) \sum_{K=1}^{\infty} \frac{\cos( \tau \nu z )}{\nu}
 \nonumber \\
  &&+
   \frac{\bar{m}^{2}}{8\pi R} \sin^{2}F \lim_{\tau \rightarrow 0_{+}} \int_{0}^{\infty} dz
           \left( -\frac{1}{h^{2}} + \frac{6}{h^{4}} - \frac{4}{h^{6}} \right) \sum_{K=1}^{\infty}
               \cos( \tau \nu z )
 \nonumber \\
   &&+ \frac{\bar{m}^{2}}{8\pi R} \sin^{2}F \lim_{\tau \rightarrow 0_{+}} \int_{0}^{\infty} dz
           \left( -\frac{1}{h^{3}} + \frac{23}{h^{5}} - \frac{41}{h^{7}} + \frac{21}{h^{9}} \right)
                \sum_{K=1}^{\infty} \frac{\cos( \tau \nu z )}{\nu}.
\end{eqnarray}
The summation over $K \ge 1$ and the integral is achieved by the formula Eq.~(\ref{eq : sum}).
Finally, we arrive at the result
\begin{eqnarray}
  \Delta E_{C}^{(K \ge 1)}(m, F) 
  &\simeq& \left[ -\frac{3}{16 R} + \left( - \frac{2}{15\pi R} \ln \tau + \frac{12\ln 2 - 9}{30\pi R} \right)         
\right.
 \nonumber \\
 &&+  \left.  \frac{\bar{m}^{2}}{64R} - \frac{1}{15} \frac{\bar{m}^{2}}{8\pi R} \left( \ln \tau + 1 -4\ln2 \right) \right] \sin^{2}F
\end{eqnarray}
Therefore, we obtain a logarithmic divergence term with coefficient of $\sin^{2}F$ for a finite quark mass.
We note that our result reproduces the previous result for the massless quarks in $\bar{m} \rightarrow 0$ \cite{Zahed84, Zahed}. 

\begin{figure}[ptb]
\begin{center}
\includegraphics[width=8cm, angle=0, clip]{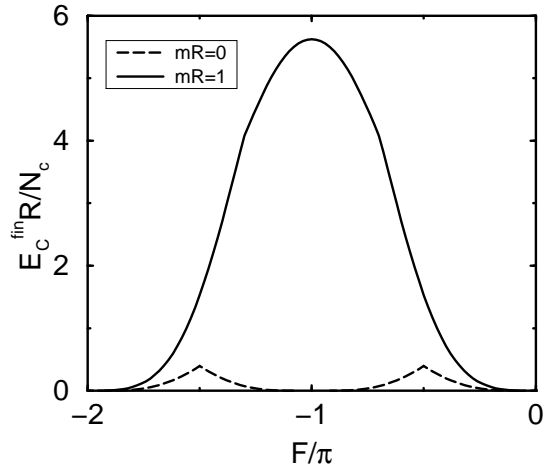}
\end{center}
\vspace*{-1.0cm} \caption{\small \baselineskip=0.5cm The Casimir energy as a function of the chiral angle at the bag surface. The dashed and solid lines indicate the quark masses $mR=0$ and $mR=1$, respectively.}
 \label{fig : Ec_F_Nmax20_mR}
\end{figure}

Now, in order to remove the logarithmic divergence, we subtract the second derivative of the chiral Casimir energy with respect to the chiral angle at the reference point $F=0$, and we obtain a finite contribution
\begin{eqnarray}
  E_{C}^{fin}(m, F) =  \Delta E_{C}(m, F) - \frac{1}{2} \sin^{2}F \left. \frac{\partial^{2}E_{C}(m, F)}{\partial F^{2}} \right|_{F=0}.
    \label{eq : chiral_Casimir_energy3}
\end{eqnarray}
In Fig.~\ref{fig : Ec_F_Nmax20_mR}, we show the chiral Casimir energy $E_{C}^{fin}$ as a function of the chiral angle $F$.
The lines are distinguished by the quark masses  $mR=0$ (dashed line), $mR=1$ (solid line), respectively.
It is a remarkable point that the Casimir energy for the massive quark takes a nonzero value at $F=-\pi$, while that of the massless quark becomes zero at $F=-\pi$.
This is because that the massive quark has an asymmetric energy spectrum between $F=0$ and $-\pi$, while the massless quark has a  symmetric spectrum, as shown in Fig.~1 in \cite{Yasui3}.

Here, we recall the Cheshire Cat picture in the chiral bag model with massless quarks.
There, the continuous transformation from the chiral bag to the Skyrmion was induced in the limit of small bag radius \cite{Brown_Rho, Brown_Rho_Vento, Vento}.
This picture seems not to be applied to the case of the massive quarks, since the chiral Casimir energy takes a finite value at the zero bag radius, or $F=-\pi$.
However, this observation does not make us abandon our discussion.
Indeed, in our formulation, the quark mass is a dynamical variable which should be determined by the energy variation.
Our previous result suggested that the chiral symmetry was restored in the limit of small bag radius \cite{Yasui3}.
Therefore, the NJL chiral bag model holds the Cheshire Cat picture.

\section{Summary}

We discussed the chiral properties in the chiral bag model which contained massive quarks.
Considering that the dynamical quark mass was generated by chiral symmetry breaking, the fractional baryon number and the chiral Casimir energy were investigated.
We discussed the effects of the finite quark mass in the hedgehog configuration with an emphasis on the technical procedure.

We showed the numerical calculation by the Strutinsky's smearing method and the analytical technique by the Debye expansion.
It was shown numerically and analytically that the fractional baryon number carried by vacuum massive quarks inside the bag was canceled that of the $\pi$ meson outside the bag.
Therefore, the total baryon number is exactly conserved.
By using the Debye expansion, it was shown that the chiral Casimir energy had a logarithmic divergence term.
The chiral Casimir energy was obtained numerically by  the Strutinsky's smearing method by removing the divergence term.
It is a point different from massless quark that the chiral Casimir energy for massive quark had a nonzero value at the chiral angle $F=-\pi$.

As further development, we plan to discuss a fully chiral symmetric equation of motion for a finite quark mass.
In our present analysis of the NJL chiral bag model, we performed a mean field approximation only in the scalar channel in the four point interaction \cite{Yasui3}.
However, in the hedgehog configuration, it may be allowed to have a finite expectation value of $\bar{q}q$, not only in the sigma channel, but also in the pion channel.
In the latter case, the expectation value of $\bar{q}q$ with pion quantum number can be given as a finite value in a basis set of the hedgehog quark wave function.
This study is now in progress.
The quantization of the hedgehog configuration for massive quark is an important subject in order to obtain the state with definite spin and isospin.
It is also an interesting subject to include the strangeness sector in our framework.
These subjects are left as future works.

\section*{Acknowledgement}
The author acknowledge to Prof.~H.~Hosaka and Prof.~M.~Oka for useful discussions.
The author is also grateful to Prof.~T.~Kunihiro.
Part of this study was proceeded when the author was belonging to
Yukawa Institute for Theoretical Physics, Kyoto University.
This work was also partially supported by Grant-in-Aid for Scientific Research for Priority Areas, MEXT (Ministry of Education, Culture, Sports, Science and Technology) with No. 17070002.


\end{document}